\documentclass[prb,twocolumn,amssymb,floatfix,amsmath,showpacs,superscriptaddress,epsfig,bm]{revtex4-1}
\usepackage{graphicx,color}
\usepackage{gensymb}

\begin{document}
\title{Investigation of Superconducting Gap Structure in HfIrSi using muon spin relaxation/rotation}
\author{A. Bhattacharyya}
\email{amitava.bhattacharyya@rkmvu.ac.in}
\affiliation{Department of Physics, Ramakrishna Mission Vivekananda Educational and Research Institute, Belur Math, Howrah 711202, West Bengal, India} 
\author{K. Panda}
\affiliation {Department of Physics, Ramakrishna Mission Vivekananda Educational and Research Institute, Belur Math, Howrah 711202, West Bengal, India}
\author{D. T. Adroja} 
\affiliation{ISIS Facility, Rutherford Appleton Laboratory, Chilton, Didcot, Oxon, OX11 0QX, United Kingdom} 
\affiliation{Highly Correlated Matter Research Group, Physics Department, University of Johannesburg, Auckland Park 2006, South Africa}
\author{N. Kase}
\affiliation{Department of Applied Physics, Tokyo University of Science, Tokyo, 125-8585, Japan}
\author{P. K. Biswas}
\affiliation{ISIS Facility, Rutherford Appleton Laboratory, Chilton, Didcot, Oxon, OX11 0QX, United Kingdom}
\author{Surabhi Saha} 
\affiliation{Department of Physics Indian Institute of Science Bangalore 560012 India}
\author{Tanmoy Das} 
\email{tnmydas@gmail.com}
\affiliation{Department of Physics Indian Institute of Science Bangalore 560012 India}
\author{M. R. Lees}
\affiliation{Department of Physics, University of Warwick, Coventry CV4 7AL, United Kingdom}
\author{A. D. Hillier}
\affiliation{ISIS Facility, Rutherford Appleton Laboratory, Chilton, Didcot, Oxon, OX11 0QX, United Kingdom} 

\begin{abstract}

Appearance of strong spin-orbit coupling (SOC) is apparent in ternary equiatomic compounds with 5$d$-electrons due to the large atomic radii of transition metals. SOC plays a significant role in the emergence of unconventional superconductivity. Here we examined the superconducting state of HfIrSi using magnetization, specific heat, zero and transverse-field (ZF/TF) muon spin relaxation/rotation ($\mu$SR) measurements. Superconductivity is observed at $T_\mathrm{C}$ = 3.6 K as revealed by specific heat and magnetization measurements. From the TF$-\mu$SR analysis it is clear that superfluid density well described by an isotropic BCS type $s$-wave gap structure. Furthermore, from TF$-\mu$SR data we have also estimated the superconducting carrier density $n_\mathrm{s}$ = 6.6 $\times$10$^{26}$m$^{-3}$, London penetration depth $\lambda_{L}(0)$ = 259.59 nm and effective mass  $m^{*}$ = 1.57 $m_{e}$. Our zero-field muon spin relaxation data indicate no clear sign of spontaneous internal field below $T_\mathrm{C}$, which implies that the time-reversal symmetry is preserved in HfIrSi. Theoretical investigation suggests Hf and Ir atoms hybridize strongly along the $c$-axis of the lattice, which is responsible for the strong three-dimensionality of this system which screens the Coulomb interaction. As a result despite the presence of correlated $d$-electrons in this system, the correlation effect is weakened, promoting electron-phonon coupling to gain importance.

\end{abstract}

\date{\today} 


\maketitle

\section{Introduction}

\noindent In recent years, ternary silicides and phosphides have drawn considerable attention due to their unusual superconducting properties~\cite{Barz1980, Muller1983, Seo1997, Ching2003, Keiber1984, Shirotani2001, Shirotani1998, Shirotani2000, Shirotani1999, Shirotani1995}. The discovery of high temperature  superconductivity with transition temperatures $>$10~K in the 111-based transition metal phosphides~\cite{Ching2003} spurred research on similar compounds formed by replacing one or more elements in the material. As a result the isoelectronic system of arsenides $MM'$As and phosphides $MM'$P [where $M$~= Zr or Hf and $M'$ =  Ru or Os], exhibits superconductivity at temperature as high as 15.5~K for hexagonal $h$-MoNiP~\cite{Shirotani2000}, 13~K for $h$-ZrRuP~\cite{Shirotani1993} and 12 K for $h$-ZrRuAs~\cite{Meisner1983}. Barz {\it et al.}~\cite{Barz1980} inferred a potentially enhanced density of states $N(E_\mathrm{F})$ at the Fermi level $E_{F}$ due to reduced Ru-Ru bond length in $h$-ZrRuP with respect to pure Ru. Ternary compounds $TrT'X$ ($Tr$, $T'$ = 4$d$ or 3$d$ transition elements and $X$ is a group V or group IV atom) with a Fe$_2$P-type hexagonal structure~\cite{Shirotani2000}; the transition temperature, $T_{\mathrm{C}}$, of these 111-based compounds are pretty high, in some cases $T_{\mathrm{C}}$ is more than 10~K, for example, Shirotani {\it et al.}~\cite{Shirotani1995} reported superconducting transitions above 10 K for both $h$-ZrRu$X$ ($X$~= P, As or Si) and $h$-HfRuP. Either $M$ and $X$ atoms or Ru and $X$ atoms occupy each layer of the hexagonal structure~\cite{Shirotani1991(1)}.  These atoms are located in each layer which are parallel to the $a-b$ plane and separated by a distance $c$. Shirotani {\it et al.}~\cite{Shirotani1991(1)} reported the formation of the two-dimensional triangular clusters of Ru$_\mathrm{3}$ in the $a-b$ plane.\\

\par

\noindent 111-based ternary compounds $TrT'X$ crystallizes  in two type of layered structures: (a) the hexagonal Fe$_2$P-type (space group $P\bar{\mathrm{6}}m$2),~\cite{Shirotani2001,Shirotani2000, Ching2003} and (b) the orthorhombic Co$_2$Si-type (space group $Pnma$)~\cite{Shirotani1998,Shirotani2000, Ching2003, Muller1983}. At high temperatures and pressures orthorhombic phase modifies to higher-symmetry hexagonal structure. Amongst this group of superconductors, the Co$_2$P-type $o$-MoRuP exhibits the highest superconducting transition temperature of 15.5~K ~\cite{Shirotani2000}. Higher value of density of states, $N(E_{\mathrm{F}})$) at the Fermi level, which are ruled by the Mo-4$d$ orbitals,  are directly linked with the higher $T_{\mathrm{C}}$ in $o$-MoRuP~\cite{Kita2004}. They have  estimated the value of $N(E_{\mathrm{F}})$ for $o$-MoRuP is 0.46 states/eV atom . Ching {\it et al.}~\cite{Ching2003} predicted that $h$-MoRuP may have a transition temperature above 20~K by assuming that the lattice dynamics and the electron-phonon coupling effect in $o$-MoRuP, $h$-MoRuP, $o$-ZrRuP, and $h$-ZrRuP crystals are reasonably close. Wong-Ng {\it et al.}~\cite{Wong2003} suggested that the key reasons for the difference in $T_{\mathrm{C}}$ are the different values in the density of states at the Fermi level.  As spin-orbit coupling (SOC) is directly proportional to $Z^4$, where $Z$ is the atomic number, strong spin-orbit coupling must be important in $o$-HfIrSi~\cite{Wang} due to the high atomic numbers of Hf, Ir. 

\noindent In order to understand the pairing mechanism in $o$-HfIrSi, we report a comprehensive study of $o$-HfIrSi using magnetization, heat capacity, zero and transverse field (ZF/TF) muon spin relaxation and rotation ($\mu$SR) measurements. Our recent $\mu$SR study on the transition metal based caged type R$_{5}$Rh$_{6}$Sn$_{18}$ (R = Lu, Sc, and Y)~\cite{BhattacharyyaLu5Rh6Sn18,BhattacharyyaY5Rh6Sn18,BhattacharyyaSc5Rh6Sn18} superconductors revealed superconductivity that breaks time-reversal symmetry (TRS) due to strong spin-orbit coupling, while recent work by Singh {\it et. al.}~\cite{Singh2014} has shown that a system with strong spin-orbit coupling can break TRS, even though conventional techniques do not show any clear evidence of unconventional superconductivity. Furthermore, our ZF-$\mu$SR data does not reveal any evidence for spontaneous internal fields below $T_{\mathrm{C}}$, which indicates that for $o$-HfIrSi  time-reversal symmetry is preserved in the superconducting state. The temperature dependence of the superfluid density determined by means of TF$-\mu$SR is described by an isotropic $s$-wave superconducting gap structure, which is in agreement with the heat capacity results.

\section{Experimental Details}

\noindent A polycrystalline sample of HfIrSi was prepared by melting stoichiometric amounts of high purity Hf, Ir, and Si in a water-cooled arc furnace. To obtain the target phase the as cast ingot was overturned and remelted several times. The sample was annealed in a sealed quartz tube for 168 hrs at 1273~K. The sample HfIrSi was characterized using a conventional X-ray diffractometer with Cu$K_\mathrm{\alpha}$ radiation. Magnetic susceptibility measurements were performed using magnetic property measurement system (MPMS) in a temperature range of 2.0-5.0~K in an applied magnetic fields. Specific heat measurements were performed by an adiabatic heat-pulse method down to 0.3~K using a commercial $^{3}$He refrigerator (Heliox-VL, Oxford Instruments) equipped with a superconducting magnet (Oxford Instruments).

\begin{figure*}[t]
\centering
    \includegraphics[width=0.7\linewidth]{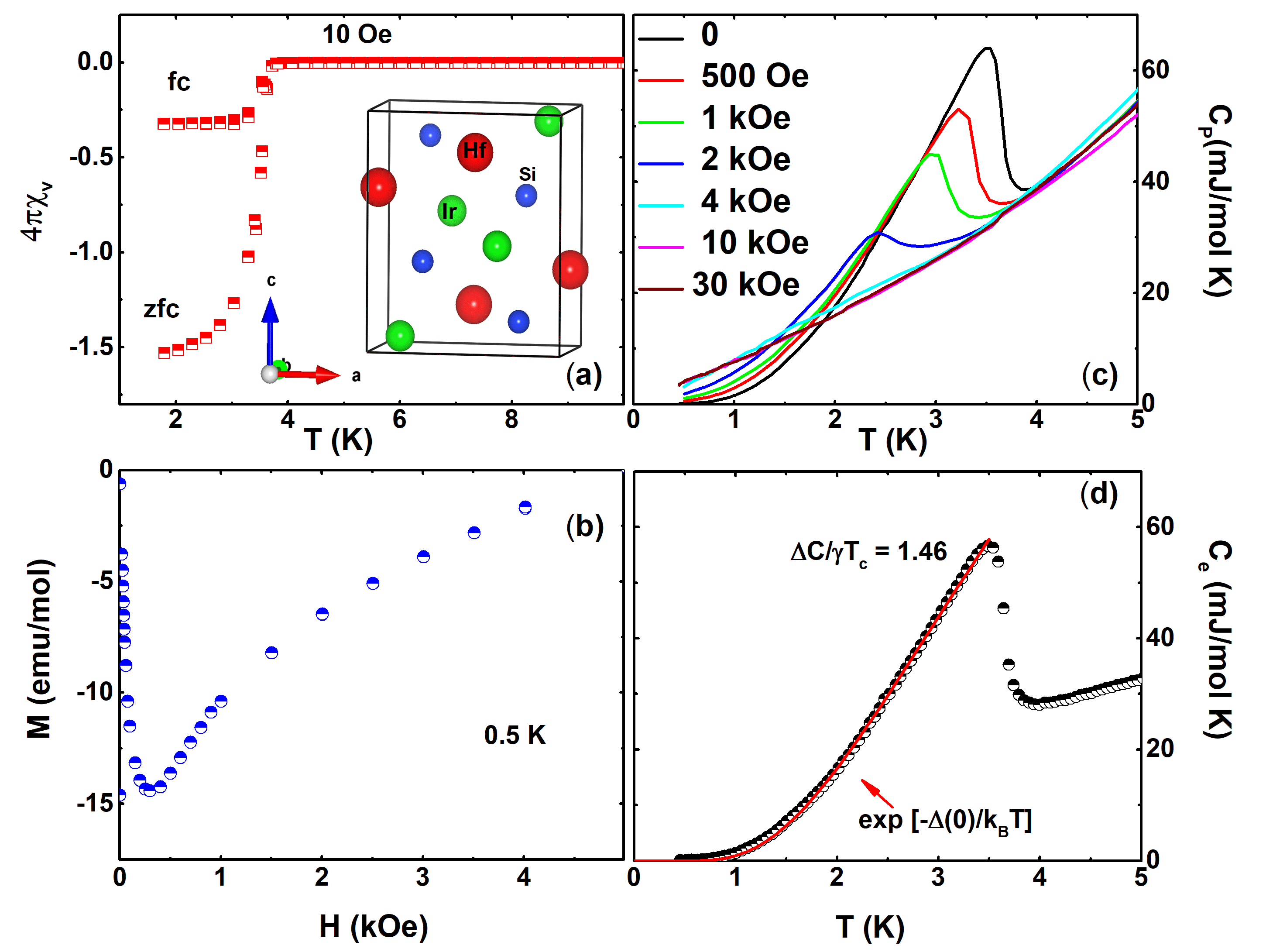}\hfil
\caption{(a) Temperature dependence of susceptibility $\chi (T)$ of HfIrSi under the field of 10 Oe for zero field cooled (ZF) and field cooled (FC) sequences. The inset shows a unit cell of orthorhombic crystal structure ($Pnma$) of HfIrSi.  (b) Isothermal field dependence of magnetization at 0.5 K. (c) Heat capacity of $C_\mathrm{P}$ of HfIrSi as a function of temperature $T$ for $0.45 \leq T \leq 5$~K measured in different applied fields. (d) Electronic contribution $C_\mathrm{e}$ to zero field heat capacity as a function of temperature. The solid line indicates the fitting.}
\label{physical properties}
\end{figure*}

\noindent The muon spin relaxation/rotation experiments ($\mu$SR) were performed on the MuSR spectrometer at the ISIS Pulsed Neutron and Muon source of the Rutherford Appleton Laboratory, UK \cite{Lee1999}. The powder sample was mounted on a silver (99.995\%) sample holder using GE-varnish, and placed in a helium-3 sorption cryostat. The time dependence of the polarization of the implanted muons is given by $P_\mathrm{\mu}(t)$ = $G(t)P_\mathrm{\mu}(0)$, where the function $G(t)$ corresponds to the $\mu^{+}$ spin autocorrelation function, which is determined by the internal magnetic field distribution~\cite{Amato}.  A commonly measured quantity in $\mu$SR experiments is the time-dependent asymmetry $A(t)$ which is proportional to  $P_\mathrm{\mu} (t)$ and is given by  $A(t)=\frac{N_{\mathrm{F}}(t)-\alpha N_{\mathrm{B}}(t)}{N_{\mathrm{F}}(t)+\alpha N_{\mathrm{B}}(t)}$, where $N_{\mathrm{F}}(t)$ and $N_{\mathrm{B}}(t)$ are the number of positrons detected in the forward and backward position respectively, and $\alpha$ is a calibration constant. Zero-field (ZF) and transverse field (TF) time dependence asymmetry spectra were collected at different temperatures between 60~mK and 4.0~K. For zero-field measuremnts, an active compensation system was used to cancel any stray magnetic fields at the sample position to a level of $\sim 0.001$~Oe. ZF-$\mu$SR is a subtle probe of the local magnetism through the precession of the muon spin in any internal magnetic fields at the muon sites. Moreover, ZF-$\mu$SR can also be used to detect the very weak spontaneous magnetic fields associated with TRS breaking in the superconducting state~\cite{Sonier}. TF-$\mu$SR measuremnts were performed in the superconducting mixed state in presence of an applied field of 300~Oe, well above the lower critical field, $H_{\mathrm{c1}} =10$~Oe, and well below the upper critical field, $H_{\mathrm{c2}}=22.3$~kOe, of HfIrSi. The main aims of the present $\mu$SR study were to examine the superconducting pairing mechanism and to search for time reversal symmetry breaking in the superconducting state of HfIrSi. All the $\mu$SR data were analyzed using WiMDA software~\cite{Pratt2000}.

\section{Results and discussion}

\subsection{Magnetization and Specific Heat}

\noindent HfIrSi crystallizes in the orthorhombic structure with space group $Pnma$ (No. 62)~\cite{Kase2016, Wang}. The lattice parameters are estimated to be $a=6.643$~\AA, $b=3.930$~\AA ~and $c=7.376$~\AA\  from the powder X-ray diffraction study. The temperature ($T$) dependence of the magnetic susceptibility $\chi$(T) of HfIrSi in an applied magnetic field of 10 Oe is shown in Fig.~\ref{physical properties}(a). $\chi(T)$ reveals a clear signature of superconductivity below the superconducting transition $T_\mathrm{C} = 3.6$~K.  The magnetization isotherm $M(H)$ shown in Fig.~\ref{physical properties}(b) at 0.5~K is typical of type II superconductivity. The lower critical field was estimated from the $M-H$ curve at 0.5~K as 10~Oe. From the field dependence of the resistivity data~\cite{Kase2016} the upper critical field was estimated to be 22.3~kOe while the Pauli paramagnetic limit 18.4$T_\mathrm{C} = 62.5$~kOe. Heat capacity ($C_{\mathrm{P}}$) as a function of temperature for $0.45 \leq T \leq 5$~K is shown in Fig.~\ref{physical properties}(c) at different applied magnetic fields. The normal state $C_{\mathrm P}$ was found to be independent of the external magnetic field. Above $T_{\mathrm{C}}$  in the normal state,  $C_\mathrm{P}$ data described using $C_\mathrm{P}(T)/T = \gamma + \beta T^{2}$, where $\gamma $ is the electronic heat capacity coefficient, and $\beta T^{3}$  is the lattice contribution. Fitting gives $\gamma = 5.56$~mJ mol $^{-1}$K$^{-2}$ and $\beta = 0.172$~mJ mol$^{-1}$K$^{-4}$. From the Debye model  $\beta$ is related to the Debye temperature ($\Theta_{\mathrm{D}}) = (\frac{12\pi^{4}}{5\beta}nN_{\mathrm{A}}R)^{1/3}$, where $R = 8.314$~J mol$^{-1}$K$^{-1}$ is the gas constant and $n = 3$ is the number of atoms per formula unit in HfIrSi. From this relationship $\Theta_{\mathrm{D}}$ is estimated to be 323~K. The jump in the heat capacity $\Delta C_\mathrm{P}(T_{\mathrm{C}}) = 28.46$~mJ/(mol-K) and $T_\mathrm{C} = 3.6$~ K, yields $\Delta C/\gamma T_\mathrm{C} = 1.42$~\cite{Kase2016}. This value is close to 1.43 expected for the weak-coupling BCS superconductors~\cite{Tinkham}.

\par

Fig.~\ref{physical properties}(d) shows the temperature dependence of the electronic specific heat $C_{\mathrm{e}}$, obtained by subtracting the phonon contribution from $C_{\mathrm P}$. $C_{\mathrm{e}}$ used to investigate the superconducting gap symmetry. From the exponential dependence of $C_{\mathrm{e}}$ we obtained $\Delta (0) = 0.50$~meV which is close to 0.51~meV, obtained from TF-$\mu$SR analysis. $\Delta (0) = 0.50$~meV gives 2$\Delta(0)/k_\mathrm{B}T_\mathrm{C} = 3.34$, which is close to the value of 3.53 expected for weak-coupling BCS superconductors~\cite{BCS}. 

\begin{figure*}[t]
\centering
    \includegraphics[width=0.6\linewidth]{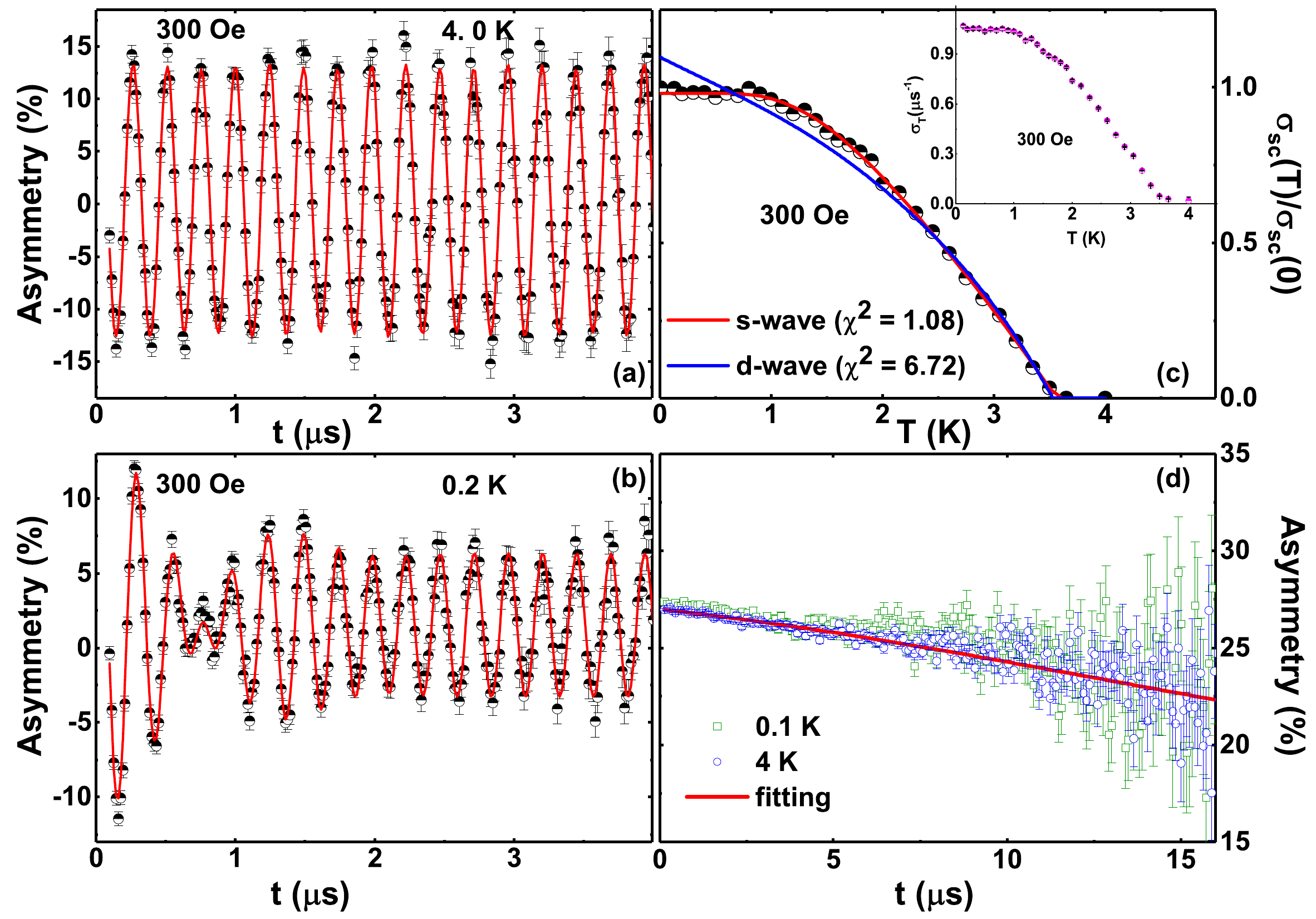}
    \caption{Transverse field $\mu$SR asymmetry spectra for HfIrSi collected (a) at $T$ = 4.0 K (b) at $T$ = 0.2 K in a magnetic field $H$ = 300 Oe. The line is a fit to the data using equation 1 in the text. (c) The temperature dependence of the superconducting depolarization rate $\sigma_\mathrm{sc}$(T) in presence of an applied magnetic field of 300 Oe. The inset shows the temperature variation of the total superconducting depolarization rate $\sigma$(T). (d) Zero field $\mu$SR time spectra for HfIrSi collected at 0.1 K (green square) and 4 K (blue circle) are shown together with lines that are least square fit to delta.}
\label{musrdata}
\end{figure*}

\subsection{Superconducting Gap Structure}

\noindent In order to investigate the nature of the pairing mechanism and the superconducting gap structure of HfIrSi compound, we performed ZF- and TF-$\mu$SR measurements. The TF-$\mu$SR asymmetry spectra above and below $T_{\mathrm{C}}$ are shown in the Figs.~\ref{musrdata}(a)-(b). Below $T_\mathrm{C}$ the asymmetry spectra decays with time due to the inhomogeneous field distribution of the flux-line lattice. The time dependence of the TF-$\mu$SR asymmetry spectra at all temperatures above and below $T_{\mathrm{C}}$, could be best fit using an oscillatory decaying muon spin depolarization function~\cite{Bhattacharyyarev, AdrojaThFeAsN, BhattacharyyaThCoC2, AnandLaIrSi3}:

\begin{equation}
G_\mathrm{z1}(t) = A_\mathrm{1}\cos(\omega_\mathrm{1}t+\phi)\exp(-\frac{\sigma^{2}t^{2}}{2})+A_\mathrm{2}\cos(\omega_\mathrm{2}t+\phi)
\end{equation}

\noindent where $A_{\mathrm{1}}$= 64.20\% and $A_{\mathrm{2}}$ = 35.80\% are the transverse field asymmetries and $\omega_{\mathrm{1}}$ and $\omega_{\mathrm{2}}$ are the frequencies of the muon precession in the sample and in the sample holder (i.e. background), respectively, $\phi$ is the initial phase of the offset and $\sigma$ is the total Gaussian muon spin relaxation rate. The superconducting contribution to the muon spin relaxation rate $\sigma_\mathrm{sc}$ is calculated using [$\sigma_\mathrm{sc} = \sqrt{\sigma^{2}-\sigma_\mathrm{n}^2}$], where $\sigma_{\mathrm{n}}$ (0.029 $\mu s^{-1}$) is the normal state contribution which is assumed to be constant over the entire temperature range and was obtained from spectra measured above $T_\mathrm{C}$. In the fitting of TF$-\mu$SR data shown in Fig.~\ref{musrdata}(a)-(b), $A_\mathrm{2}$ was fixed at the value of 0.358 obtained from fitting the lowest temperature data set. $A_\mathrm{1}$ was allowed to vary, but its value of 0.672 is nearly temperature independent. The value of the phase was also estimated from the lowest temperature data set and then fixed for all other temperatures. The temperature dependence of the penetration depth/superfluid density modelled using~\cite{Prozorov, BhattacharyyaBiS2, AdrojaK2Cr3As3, AdrojaCs2Cr3As3}

\begin{align}
\frac{\sigma_\mathrm{sc}(T)}{\sigma_\mathrm{sc}(0)} &= \frac{\lambda^{-2}(T,\Delta_\mathrm{0,i})}{\lambda^{-2}(0,\Delta_\mathrm{0,i})}\\ \nonumber
 &= 1 + \frac{1}{\pi}\int_{0}^{2\pi}\int_\mathrm{\Delta(T)}^{\infty}(\frac{\delta f}{\delta E}) \times \frac{EdEd\phi}{\sqrt{E^{2}-\Delta(T,\Delta_{i}})^2} 
\end{align}

\noindent where $f= [1+\exp(-E/k_\mathrm{B}T)]^{-1}$ is the Fermi function and $\Delta_\mathrm{i}(T,0) = \Delta_\mathrm{0,i}\delta(T/T_\mathrm{C})\mathrm{g}(\phi)$. The temperature variation of the superconducting gap is approximated by the relation $\delta(T/T_\mathrm{C}) = \tanh[1.82[1.018(T_\mathrm{C}/T-1)]^{0.51}]$ where $\mathrm{g}(\phi$) refers to the angular dependence of the superconducting gap function and $\phi$ is the polar angle for the anisotropy. $\mathrm{g}(\phi$) is replaced by (a) 1 for an $s$-wave gap, and (b) $\vert\cos(2\phi)\vert$ for a $d$-wave gap with line nodes\cite{Pang2015, Annet1990}. The data best modeled using a single isotropic $s$-wave gap of 0.51~meV. This gives a gap of 2$\Delta/k_\mathrm{B}T_\mathrm{C} = 3.38$, which is very close to the gap value 3.34 obtained from the heat capacity data presented earlier, and indicating a weak-coupling superconductivity in HfIrSi. 

\begin{figure}[b]
\centering
    \includegraphics[width=\linewidth]{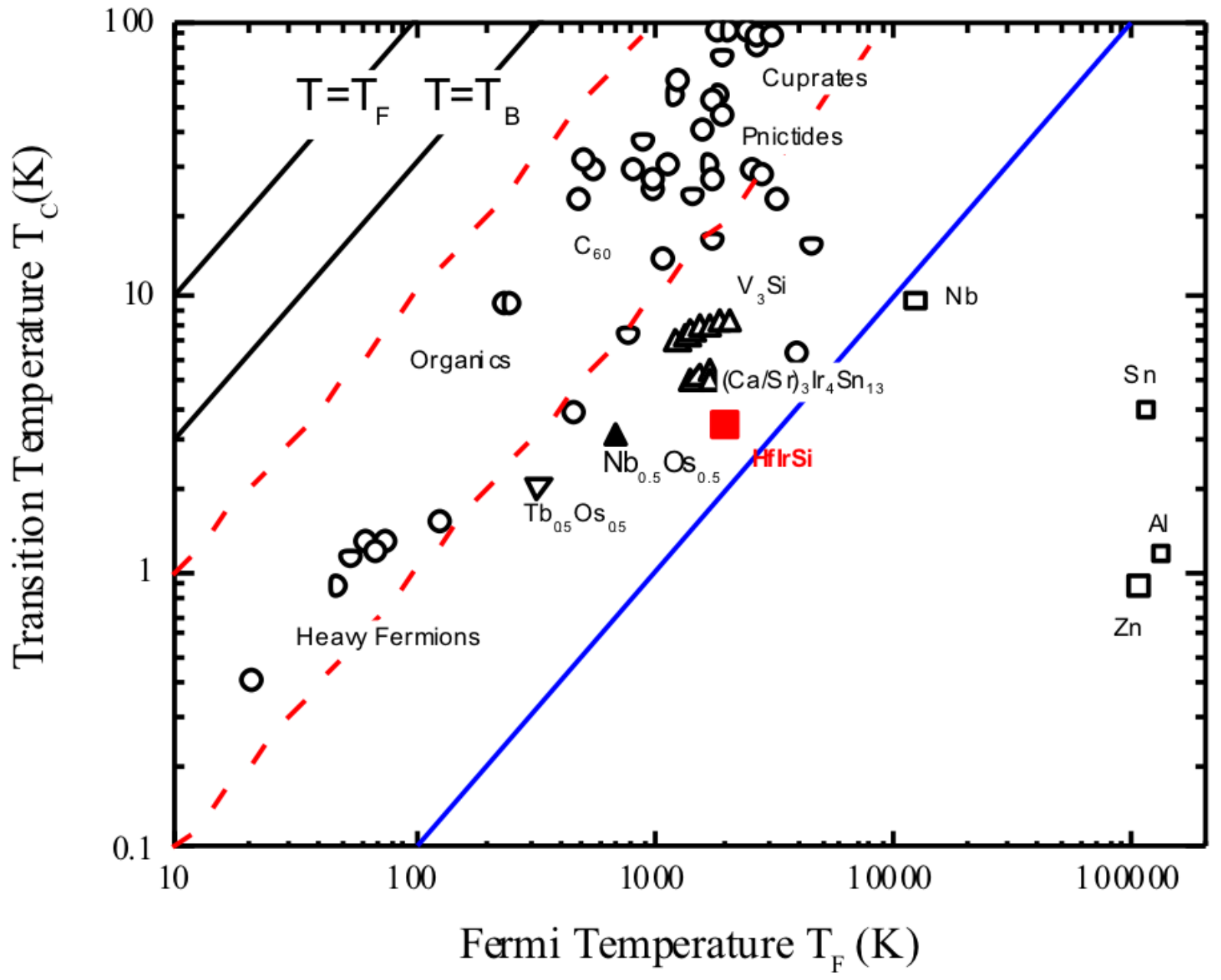}
\caption{ (a) Uemura plot of $T_\mathrm{C}$ vs. effective Fermi temperature $T_\mathrm{F}$. The ``exotic" superconductors lie within a region bounded by 1/100$\le T_\mathrm{C}$/$T_\mathrm{F}\le$1/10, as indicated by the red dashed lines. The solid black line correspond to the Bose-Einstein condensation temperature ($T_\mathrm{B}$). The positions of HfIrSi on the plot place these materials as being conventional superconductors.}
\label{Uemura plot}
\end{figure}

\subsection{Superconducting Parameters}

\noindent The muon spin depolarization rate observed below the superconducting transition temperature is related to the superfluid density or penetration depth. For a triangular\cite{EHB, Sonier,Chia,Amato} lattice  $\frac{\sigma_\mathrm{sc}^2}{\gamma_\mathrm{\mu}^2}=\frac{0.00371 \times \phi_\mathrm{0}^{2}}{\lambda^4}$, where $\phi_{\mathrm{0}}$ is the flux quantum number (2.07 $\times$10$^{-15}$Tm$^{2}$) and $\gamma_{\mathrm{\mu}}$ is the muon gyromagnetic ratio $\gamma_\mathrm{{\mu}}/2\pi$ = 135.5 MHz T$^{-1}$. As with other phenomenological parameters characterizing the superconducting state, the superfluid density can also be related to quantities at the atomic level. Using London's theory \cite{Sonier} $\lambda_{\mathrm{L}}^2 = \frac{m^{*}c^{2}}{4\pi n_\mathrm{s}e^{2}}$, where $m^{*} = (1+\lambda_\mathrm{e-ph})m_\mathrm{e}$ is the effective mass and $n_\mathrm{s}$ is the density of superconducting carriers. Within this simple picture, $\lambda_{\mathrm{L}}$ is independent of magnetic field. $\lambda_{\mathrm{e-ph}}$ is the electron-photon coupling constant that estimated from the Debye temperature ($\Theta_{\mathrm{D}}$) and $T_{\mathrm{C}}$ using McMillans relation \cite{McMillan,BhattacharyyaLaIr3, DasLaPt2Si2}

\begin{equation}
\lambda_\mathrm{e-ph} = \frac{1.04+\mu^{*}\ln(\Theta_\mathrm{D}/1.45T_\mathrm{C})}{(1-0.62\mu^{*})\ln(\Theta_\mathrm{D}/1.45T_\mathrm{C})-1.04}
\end{equation}

\noindent Here $\mu^{*}$ is the repulsive screened Coulomb parameter with a typical value of $\mu^{*}$ = 0.13, give $\lambda_{\mathrm{e-ph}} = 0.5685$. As HfIrSi is a type II superconductor, supposing that approximately all the normal states carriers ($n_\mathrm{e}$) contribute to the superconductivity ($n_\mathrm{s} \approx n_\mathrm{e}$), the magnetic penetration depth $\lambda$, superconducting carrier density $n_\mathrm{s}$, the effective-mass enhancement $m^{*}$ have been estimated to be $\lambda_\mathrm{L}(0)$ = 259.59~nm, $n_\mathrm{s} = 6.605 \times 10^{26}$ carriers m$^{-3}$, and $m^{*} = 1.577m_\mathrm{e}$ respectively, for HfIrSi.

\subsection{Time Reversal Symmetry}

\noindent ZF$-\mu$SR  were used to check for the presence of any hidden magnetic ordering in HfIrSi. Fig.~\ref{musrdata}(d) compares the zero field time-dependent asymmetry spectra above and below and $T_\mathrm{C}$ (for $T = 0.1$ and 4.0~K). The spectra are found to be identical. There is no sign of muon spin precession, ruling out the presence of any large internal magnetic fields as seen in magnetically ordered compounds. The ZF$-\mu$SR data well described using a damped Gaussian Kubo-Toyabe (KT) function, 

\begin{equation}
G_\mathrm{z2}(t) = A_\mathrm{3}G_\mathrm{KT}(t)e^{-\lambda_\mathrm{\mu}t}+A_\mathrm{bg},
\label{GKTfunction}
\end{equation}

where $G_\mathrm{KT}(t) = [\frac{1}{3}+\frac{2}{3}(1-\sigma_\mathrm{KT}^{2}t^{2})\exp({-\frac{\sigma_\mathrm{KT}^2t^2}{2}})]$, is known as the Gaussian Kubo-Toyabe function, $A_\mathrm{3}$ is the zero field asymmetry of the sample signal, $A_\mathrm{bg}$ is the background signal, $\sigma_\mathrm{KT}$ and $\lambda_\mathrm{\mu}$ are the muon spin relaxation rates due to randomly oriented nuclear moments (the local field distribution width $H_\mathrm{\mu} = \sigma/\gamma_\mathrm{\mu}$, with muon gyromagnetic ratio $\gamma_\mathrm{\mu}/2\pi$ = 135.53 MHz/T. The fitting parameters $A_\mathrm{3}$, $A_\mathrm{bg}$ and $\sigma_\mathrm{KT}$ are found to be temperature independent. There is no noticeable change between the relaxation rates at 4.0~K ($\geq T_\mathrm{C}$) and 0.1~K ($\leq T_\mathrm{C}$). Fits to the ZF$-\mu$SR asymmetry data using Eq.~\ref{GKTfunction} and shown by the solid lines in Fig.~\ref{musrdata}(d) give $\sigma_\mathrm{KT} = 0.065 ~\mu \mathrm{s}^{-1}$ and $\lambda_\mathrm{\mu} = 0.0046 ~\mu \mathrm{s}^{-1}$ at  0.1~K and $\sigma_\mathrm{KT} = 0.069 ~\mu \mathrm{s}^{-1}$ and $\lambda_\mathrm{\mu} = 0.0046 ~\mu \mathrm{s}^{-1}$ at 4~K. The values of both $\sigma_\mathrm{KT}$ and $\lambda_\mathrm{\mu}$ at $T\leq T_\mathrm{C}$ and $ \geq T_\mathrm{C}$ agree within error, indicating that time reversal symmetry is preserved in the superconducting state in HfIrSi.

\subsection{Uemura Classification Scheme}

\noindent In this section we focus upon the Uemura classification scheme~\cite{Uemura, Hillier1} which is based on the correlation between the superconducting $T_\mathrm{C}$ and the effective Fermi temperature, $T_\mathrm{F}$ ( = $\frac{\hbar^{2}(3\pi^{2})^{2/3}n_\mathrm{s}^{2/3}}{2k_\mathrm{B}m^{*}}$), determined from $\mu$SR measurements of the superconducting penetration depth. Within this scheme strongly correlated ``exotic" superconductors, i.e. high $T_\mathrm{C}$ cuprates, heavy fermions and organic materials, all lie within a particular part of the diagram, which is indicative of a degree of universal scaling~\cite{Hillier1} of $T_\mathrm{C}$ with $T_\mathrm{F}$ such that $1/10\geq (T_\mathrm{C} /T_\mathrm{F}) \ge 1/100$. For conventional BCS superconductors $1/1000 \geq T_\mathrm{C} / T_\mathrm{F}$.  The position of HfIrSi ($T_\mathrm{C}/T_\mathrm{F}$ = 3.4/2316.6 = 0.00167) on the plot places this material as being conventional superconductor. 

\subsection{Theoretical Investigations}
\noindent HfIrSi belongs to the $Pnma$($62$) space group and possess the orthorhombic  crystal system with the $mmm$ point group symmetry. We use Vienna Ab-initio Simulation Package (VASP) for density-functional theory (DFT) electronic structure calculations. The projected augmented wave (PAW) pseudo-potentials are used to describe the core electrons, and Perdew-Burke-Ernzerhof (PBE) functional is used for the exchange-correlation potential. Cut-off energy for plane wave basis set is used to be 500~eV. The Monkhorst-Pack $k$-mesh is set to $14\times14\times14$ in the Brillouin zone for the self-consistent calculation. We obtained the relaxed lattice parameters as $a=3.944\AA$, $b=6.499\AA$, $c=7.376\AA$ and $\alpha=\beta=\gamma=90\degree$. To deal with the strong correlation effect of the $d$-electrons of the Ir atoms, we employed LDA+U method with $U=2.8$~eV. For the Fermi surface and density of states calculations, we have used a larger $k$-mesh of $31\times31\times31$. We repeat the calculation with SOC and find no considerable change in the low-energy spectrum.
 
 \begin{figure}[t]
\centering
\includegraphics[width=\linewidth]{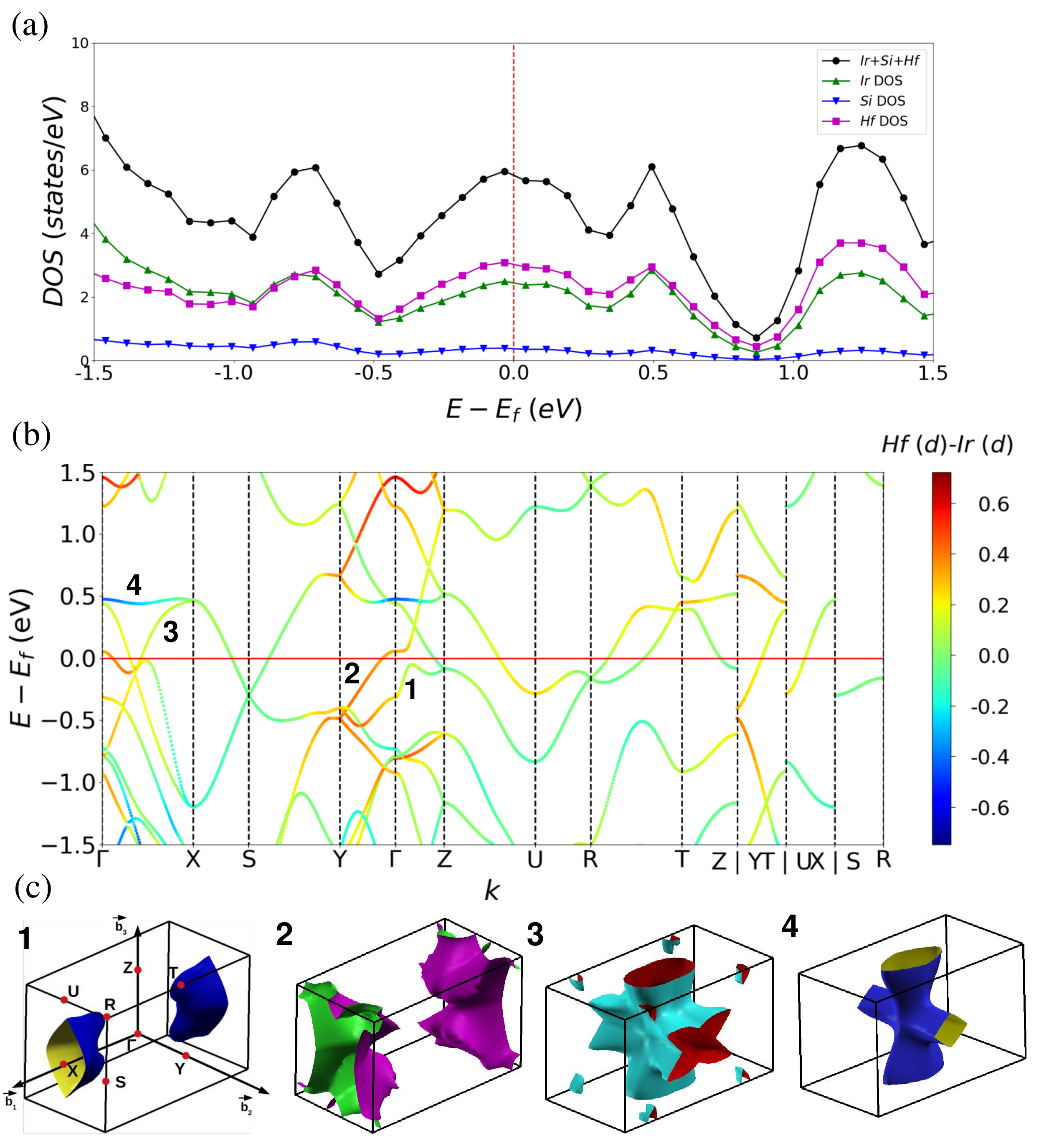}
\caption{(a) Computed partial DOSs for the three atoms are compared along with the total DOS. We clearly observe that Hf and Ir atoms contribute dominantly to the low-energy spectrum. For both these atoms, $5d$ and $4d$ orbitals contribute mostly in this energy scale. (b) DFT result of band structure is plotted along the standard high-symmetric directions for the orthorhombic crystal structure of HfIrSi. The band structure is colored with a gradient colormap with blue color means Ir-$d$ orbitals and red color means Hf-$d$ orbitals. Four Fermi surfaces are denoted by `1', `2', `3', and `4' numbers. The result indicates that the bands near the $\Gamma$-point are dominated by Hf-$d$ states while those near the  `Z' point gain more Ir-$d$ weights. (c) Corresponding four Fermi surfaces in the 3D Brillouin zone. The color on the Fermi surface has no significance here.}
\label{DFT}
\end{figure}

Due to the involvement of the strongly correlated transition metals as well as possible SOC, one may anticipate superconductivity to be exotic in this material. However, the observation of conventional, time-reversal invariant superconductivity leads to an essential question: how does the phonon mediated attractive potential win over the strong Coulomb interaction to form conventional superconductivity. 

To address this point, we investigate the DFT band structure as shown in Fig.~\ref{DFT}. From the partial-DOS result in Fig.~\ref{DFT}(a), we find that both transition metals Ir and Hf have nearly equal weight in the low-energy spectrum, and contribute mostly to the total DOS. Additionally, both transition metals have the corresponding $4d$ and $4d$ orbitals, respectively providing dominant contribution to the Fermi surfaces. The result indicates that these two atoms have strong hybridization in this system. This is confirmed by the visualization of the orbital weight distributions on the electronic structure as shown in Fig.~\ref{DFT}(b). We plot here the difference in the orbital weights between the Ir-$d$ and Hf-$d$ orbitals. We indeed find that the bands near the $\Gamma$-point are dominated by the Hf-$d$ orbitals while those on near the `Z' (on the $k_z=\pm \pi$-plane) are contributed strongly by the Ir-$d$  atoms. This result indicates that the Hf and Ir atoms hybridize rather strongly along the $c$-axis of the lattice. This hybridization is responsible for the strong three-dimensionality of this system which screens the Coulomb interaction. As a result despite the presence of correlated $d$ electrons in this systems, the correlation effect is weakened, promoting electron-phonon coupling to gain importance.\cite{EP}

The Fermi surface topologies as shown in Fig.~\ref{DFT}(c) confirm the strong three dimensionality in all four bands. Consistently we find two Fermi pockets near the $\Gamma$-points and two Fermi pockets centering the `X'-points. The large Fermi surface volume is consistent with the large carrier density of this system as measured in our $\mu$SR experiments, and higher values of Fermi temperature as presented in Fig.~3. These results are consistent with the weak correlation strength in this system.\cite{DasAIP} Strong three-dimensionality can also reduce the relativistic effect and hence the SOC strength is reduced.

\section{COMMENTS AND PERSPECTIVES}

We have presented the magnetization, heat capacity, ZF and TF$-\mu$SR measurements in the normal and superconducting states of HfIrSi, which crystallizes in the orthorhombic crystal structure. Our magnetization and heat-capacity measurements confirmed bulk superconductivity in this material with $T_{\mathrm{C}} = 3.6$~K. From the TF$-\mu$SR we have determined the muon depolarization rate in the field cooled mode associated with the vortex lattice. The temperature dependence of $\sigma_{\mathrm{sc}}$ is better fit by an isotropic $s$-wave gap model than a $d$-wave model. The value of 2$\Delta(0)/k_\mathrm{B}T_{\mathrm{C}} = 3.38$ obtained from the $s-$wave gap fit, suggests a weak-coupling BCS type superconductivity in HfIrSi. The ZF$-\mu$SR measurements reveal no sign of spontaneous field appearing below the superconducting transition temperature which suggests that time-reversal symmetry is preserved below $T_\mathrm{C}$. Theoretical investigation suggests Hf and Ir atoms hybridize strongly along the $c$-axis of the lattice, which is responsible for the strong three-dimensionality of this system which screens the Coulomb interaction. As a result despite the presence of correlated $d$-electrons in this systems, the correlation effect is weakened, promoting electron-phonon coupling to gain importance. To date, a large number of equiatomic ternary metal compounds have been discovered with high superconducting transition temperatures and high critical magnetic fields, but $\mu$SR investigations have been carried out just on a few compounds. The present study will provide an invaluable comparison for the future $\mu$SR investigations on these families of compounds. The present results will help to develop a realistic theoretical model including the role of strong spin-orbital coupling to understand the origin of superconductivity in HfIrSi and also may help us to arrive at an empirical criterion for the occurrence of superconductivity with strong SOC and high $T_c$ and $H_{c2}$ in other equiatomic ternary systems.

\subsection*{Acknowledgements}{AB would like to acknowledge the Department of Science and Technology (DST) India, for an Inspire Faculty Research Grant (DST/INSPIRE/04/2015/000169), and the UK-India Newton grant for funding support. KP acknowledge the financial support from DST India, for Inspire Fellowship (IF170620). DTA would like to thank the Royal Society of London for the UK-China Newton funding and the Japan Society for the Promotion of Science for an invitation fellowship. KP would like to acknowledge the DST India, for Inspire Fellowship (IF170620). TD acknowledges the financial support from Science and Engineering Research Board (SERB), Department of Science \& Technology (DST), Govt. of India for the Start Up Research Grant (Young Scientist).}

\end{document}